# Hubble Redshift

W. Q. Sumner

*Kittitas Institute of Astrophysics, Box 588, Kittitas, WA 98934 USA*

*kia@elltel.net*

E. E. Vityaev

*Institute of Mathematics, Russian Academy of Sciences, Novosibirsk, 630090 Russia*

*vityaev@math.nsc.ru*

Recent measurements of Hubble redshift from supernovae are inconsistent with the standard theoretical model of an expanding Friedmann universe. Figure 1 shows the Hubble redshift for 37 supernovae measured by Riess *et al.*[1] illustrating that a positive cosmological constant must be added to the equations of general relativity to fit the data. A negative deceleration parameter was also required, implying that the universe is accelerating in its expansion, a surprising result since it implies the existence of a repulsive force that overpowers gravity at cosmological distances. This letter shows that a much simpler explanation of these anomalous observations follows directly from a forgotten paper written by Erwin Schrödinger[2] in 1939. Neither a cosmological constant nor repulsive matter is required to explain the Hubble redshift of supernovae. Relativistic quantum mechanics is enough.

Erwin Schrödinger[2] proved that all quantum wave functions coevolve with Friedmann spacetime geometry. The plane-wave eigenfunctions characteristic of flat spacetime are replaced in curved spacetime by eigenfunctions whose wavelengths are directly proportional to the Friedmann radius. This means that as the radius of the universe changes every eigenfunction changes wavelength and hence every quantum system changes with spacetime curvature.

Schrödinger's beautifully simple result reflects the intrinsic connection between wave solutions and boundary conditions characteristic of every quantum system. Doubling the size of the universe doubles the wavelength of every one of its eigenfunctions just as doubling the width of a square well doubles the wavelength of every one of its eigenfunctions. In an expanding universe quantum systems expand. In a contracting universe they contract. Schrödinger's result is quite general: If you accept the logic of relativistic quantum mechanics and assume the closed Friedmann spacetime metric, Schrödinger's conclusion necessarily follows. The dependence of quantum wave functions on spacetime curvature is not an assumption that you are free to make or ignore at will.

Schrödinger explained the redshift of photons in an expanding universe and also established the evolution of atoms[3]. Atomic evolution is critical to the explanation of Hubble redshift since redshift is determined by comparing starlight emitted by atoms long ago to today's reference atoms.

The Friedmann line element for the closed solution may be written

$$ds^2 = c^2 dt^2 - a^2(t)\left[\frac{dr^2}{1-r^2} + r^2\left(d\vartheta^2 + \sin^2\vartheta\, d\varphi^2\right)\right]. \tag{1}$$

*a(t)*, the Friedmann radius, is plotted in Figure 2.

As Schrödinger[2] proved, the wavelength of a quantum system $\lambda(t)$ (and consequently its momentum) is directly proportional to the radius of the Friedmann universe,

$$\frac{\lambda(t_0)}{\lambda(t_1)} = \frac{a(t_0)}{a(t_1)}. \tag{2}$$

While the energy of a photon is directly proportional to its momentum, the energy of a particle is proportional to the *square* of its momentum. The shift in wavelength of a characteristic atomic emission $\lambda_e(t)$ due to evolution is then

$$\frac{\lambda_e(t_0)}{\lambda_e(t_1)} = \frac{a^2(t_0)}{a^2(t_1)}. \tag{3}$$

Consider an excited atom, the photon it emits, and the implications of coevolution in an expanding universe. Referring to Figure 3, the wavelength of an emitted photon at a past time $t_1$ is $\lambda(t_1)$. As the universe expands, the wavelength of this photon redshifts to $\lambda(t_0)$ according to equation (2). The characteristic emission of an excited atom also redshifts but at the different rate of equation (3). When the redshifted photon is compared against the laboratory standard of the new atomic emission, a relative blueshift is observed, even though the older photon has indeed redshifted since its emission. The excited atom that provides the measuring standard has simply out redshifted the photon. *Blueshifts are characteristic of expanding Friedmann universes*[4].

An identical analysis for a contracting universe shows that both atoms and photons blueshift, but that atomic emissions out blueshift photons giving the relative redshift Hubble observed. *Redshifts are characteristic of contracting Friedmann universes*. A contracting universe must be closed, justifying Schrödinger's original assumption that it is.

Since Hubble redshift is a result of changes in both photon wavelengths and in atomic standards, it is necessary to re-derive the connection between measured redshift, the Hubble constant $H_0$ (which is negative for contracting universes), the deceleration parameter $q_0$, and the distances to the sources.

The measured redshift $k$ is

$$k = \frac{\lambda_{obs}(t_0) - \lambda_e(t_0)}{\lambda_e(t_0)}, \qquad (4)$$

where $\lambda_e(t_0)$ is the wavelength emitted by today's reference atom and $\lambda_{obs}(t_0)$ is the photon wavelength observed today from a distant source.

The mathematical coordinate distance $r$ of a light source is directly related to its observed redshift $k$ and the deceleration parameter $q_0$. The derivation for the contracting universe is essentially the same as the one for an expanding universe when atomic evolution is ignored, except that $k$, not $z$, describes the observed redshift and some choices in the signs of square roots must be made differently[5]. It is assumed that the observed photons were emitted after contraction began. $r$ for a contracting universe is given by

$$r = \frac{(2q_0 - 1)^{\frac{1}{2}}}{q_0}\left[k - \frac{(1+k)(1-q_0)}{q_0}\right] + \frac{(1-q_0)}{q_0}\left\{1 - \left[k - \frac{(1+k)(1-q_0)}{q_0}\right]^2\right\}^{\frac{1}{2}}. \qquad (5)$$

$r$ is not directly observable by astronomers but is related to the luminosity distances and relative magnitudes that are. Luminosity distance $D_L$ is connected to the measured flux $f$ and the luminosity $L$ of a source by

$$f = \frac{L}{4\pi D_L^2}. \qquad (6)$$

The relative distances to sources with the same luminosity (or "standard candles") can then be determined by measuring their relative brightness.

Here it is assumed that a standard candle is a source of photons from a known atomic transition pulsing at a constant rate as defined by the time it takes light to travel some multiple of an atom radius. The flux observed can then be calculated by noting that the luminosity is decreased by a factor of $a(t_0)/a(t_1)$ because of the apparent decrease of the photon's energy and decreased by another factor of $a(t_0)/a(t_1)$ because of changes in local time². The distance to the source is $r\,a(t_0)$. This gives an observed flux of

$$f = \frac{L \dfrac{a^2(t_0)}{a^2(t_1)}}{4\pi\, r^2 a^2(t_0)}. \qquad (7)$$

Comparing (6) and (7) gives

$$D_L = r\,a(t_0)\,(1+k). \qquad (8)$$

Combining equations (5) and (8) gives the desired result

$$D_L = \left(\frac{-c}{H_0}\right)\frac{(1+k)}{q_0}\left\{\left[k - \frac{(1+k)(1-q_0)}{q_0}\right] + \frac{(1-q_0)}{(2q_0-1)^{1/2}}\left\{1-\left[k-\frac{(1+k)(1-q_0)}{q_0}\right]^2\right\}^{1/2}\right\}. \qquad (9)$$

The relationship between measured magnitude, *m-M*, and luminosity distance, $D_L$ is

$$m - M = 5\log_{10}\left(\frac{D_L}{10\text{ parsecs}}\right). \qquad (10)$$

While astronomical objects are more complicated than the pulsing source defined here, they are still quantum systems whose wavefunctions evolve precisely in the same way Schrödinger articulated, with their cosmological redshift described by the same equations.

$H_0$ and $q_0$ were varied in equation (9) to fit the data set of Riess *et al.*[1] Figure 4 illustrates the result for the parameters $H_0 = -65\ km\ s^{-1}\ Mpc^{-1}$ and $q_0 = 0.6$. This is as good as the best fit found by Riess *et al.* but does not assume a cosmological constant or require a negative deceleration.

This demonstrates that it is straightforward to understand the measurements of Hubble redshift from supernovae. One must simply assume the validity of relativistic quantum mechanics and the Friedmann solution to the equations of general relativity and follow these assumptions to their logical conclusion.

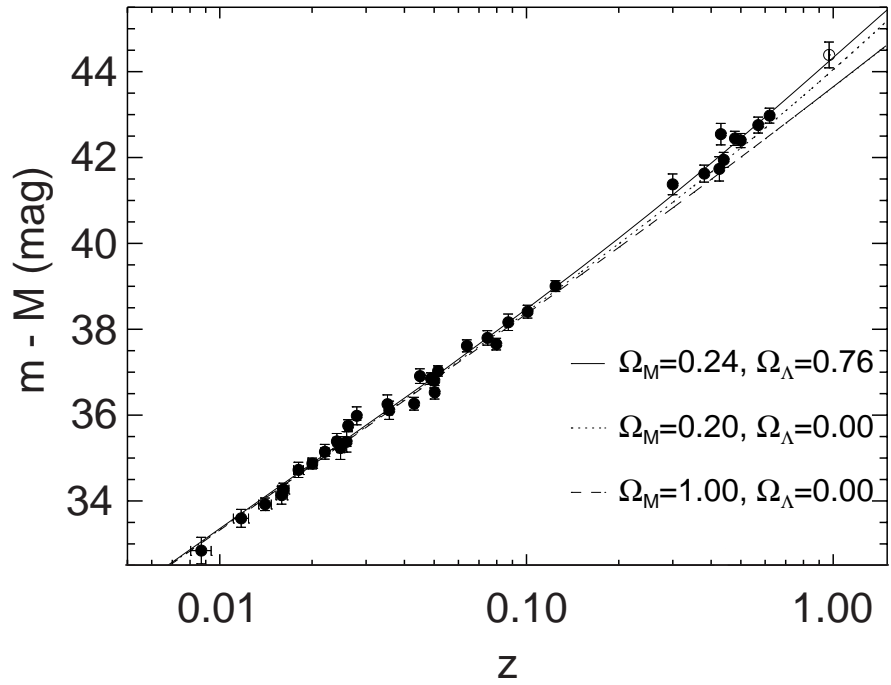

Figure 1: Redshift and magnitude for 37 supernovae and best fits for three mixes of ordinary matter $\Omega_M$ and vacuum energy resulting from a cosmological constant $\Omega_\Lambda$ (Riess *et al.*[1]).

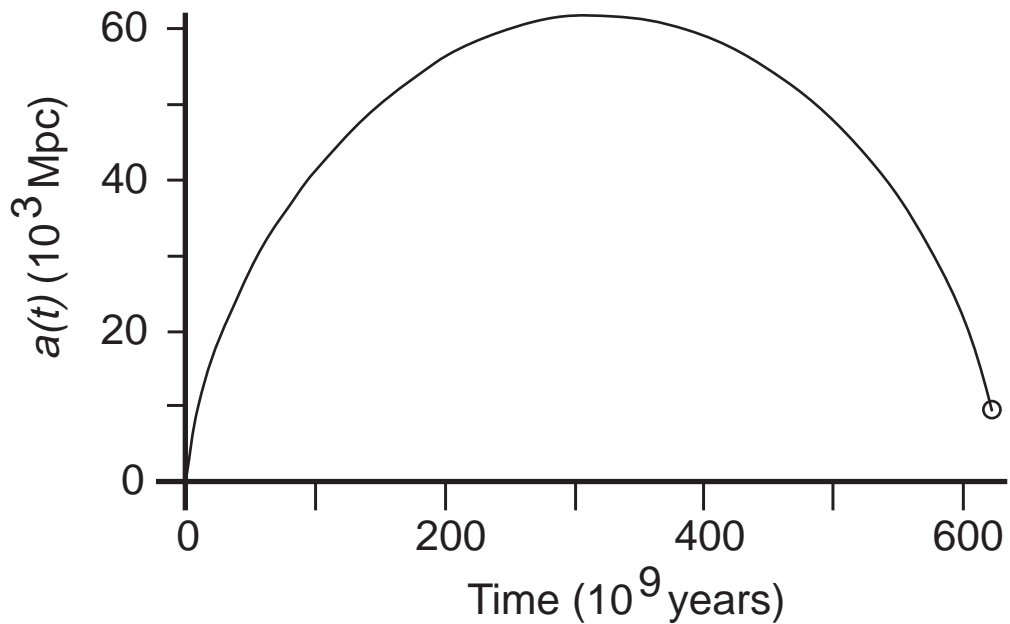

Figure 2: The Friedmann radius $a(t)$ as function of time for the parameters $H_0 = -65\ km\ s^{-1}\ Mpc^{-1}$ and $q_0 = 0.6$. The current age of the universe implied by these parameters is 623 billion years.

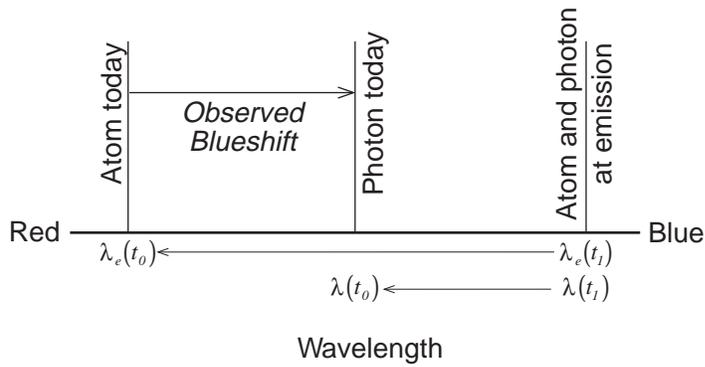

Figure 3: Shifts in wavelenths for a photon, $\lambda(t)$, and an atomic emission, $\lambda_e(t)$, from the time of emission $t_1$ to the present time $t_0$ for an expanding universe, $a(t_0) > a(t_1)$. A blueshift is observed.

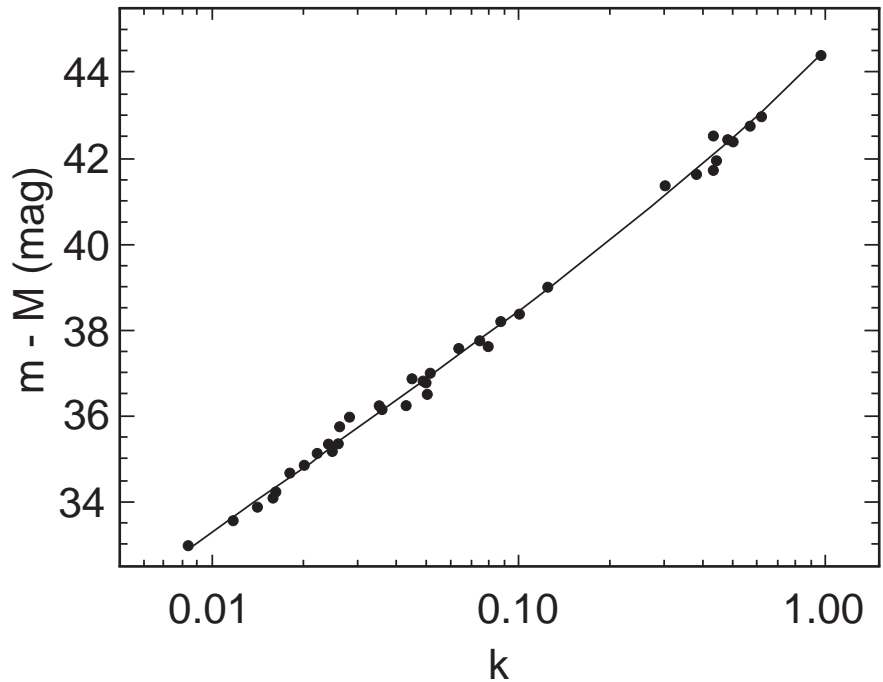

Figure 4: Redshift and magnitude for 37 supernovae (Riess *et al.*[1]) and a fit with the parameters $H_0 = -65\ km\ s^{-1}\ Mpc^{-1}$ and $q_0 = 0.6$ using equation (9).